\documentclass[12]{article}

\usepackage[utf8]{inputenc}
\usepackage{amssymb, amsmath, amscd}
\usepackage{graphicx} 
\usepackage{rotating}
\usepackage{tikz-cd}
\usepackage{float}

\usepackage{color}
\usepackage{subfig, color}

\newtheorem{lemma}{Lemma} 
\newtheorem{proposition}{Proposition} 
\newenvironment{system}[1][]%
	{\begin{eqnarray} #1 \left\{ \begin{array}{lll}}%
	{\end{array} \right. \end{eqnarray}}

\title{Enhancing the efficiency of adiabatic quantum computations}
\author{Raouf Dridi\footnote{rdridi@andrew.cmu.edu}, \, Hedayat Alghassi\footnote{halghassi@cmu.edu} ,\, Sridhar Tayur\footnote{stayur@cmu.edu} \\
\small CMU Quantum Computing Group\\
\small Tepper School of Business, Carnegie Mellon University, Pittsburgh, PA 15213\\}

\begin{document}
\maketitle

\begin{abstract}
    We describe a general methodology for enhancing the efficiency of adiabatic quantum computations (AQC). It consists of homotopically deforming  the original ``Hamiltonian surface" in a way  that the redistribution of the Gaussian curvature weakens the effect of the anti-crossing, thus yielding the desired improvement. Our approach is not pertubative but instead is built on our previous global description of AQC in the language of Morse theory. Through the homotopy deformation we witness the birth and death of critical points whilst, in parallel,  the Gauss-Bonnet theorem reshuffles the curvature around the changing set of critical points. Therefore, by creating  enough  critical points around the anti-crossing, the total curvature--which was initially centered at the original anti-crossing--gets redistributed around the new neighbouring critical points, which weakens its severity and so improves the speedup of the AQC. We illustrate this on two examples taken from the literature. 
\end{abstract}

\newpage
\section{Introduction}
An important early result in quantum mechanics is the so-called avoided crossing, also anti-crossing theorem. 
It dates back to the work of J. von Neumann, and E. Wigner~\cite{vonNeumann1993} -- more accurately,  two years  earlier, to the work of F. Hund~\cite{Hund1927};  
J. von Neumann and E. Wigner   proved  
 and generalized  
the intuition of F. Hund. 
Their finding  is as follows  (see also \cite{Suzuki2013, cmu2}): 

~~\\
Suppose the 
Hermitian matrix~$H$ depends on a number of parameters $\kappa_1, \kappa_2, \cdots$. Then {\it when one can change only
one or two parameters it is in general not possible to reach an intersection of the eigenvalues of $H$}. This number goes to one if $H$ is a real Hermitian matrix; that is,
{\it in
this case it suffices to be able to change two real parameters to make two eigenvalues degenerate. }

~~\\
The role of anti-crossings in adiabatic quantum computations is well known~\cite{Farhi472, Vazirani}. 
This is expressed as the total adiabatic evolution time $T$ being inversely proportional to the square of the energy difference between the two lowest energies of 
the given Hamiltonian 
\begin{equation}\label{H}
    H(s) = (1-s) H_{initial}  + s H_{final},
\end{equation} 
for $0\leq s=t/T\leq 1$. 
 This energy gap, for many instances\footnote{By an instance, we mean  a time dependent Hamiltonian; thus, we  include the choice of the initial Hamiltonian in our definition--and not just the problem Hamiltonian.}, is very small,  and the enlargement of this gap is the  focal point in current research in the area of adiabatic quantum computations. The goal of our paper is to present a general principled methodology to study anti-crossings that generalizes previous studies (that are ad-hoc) and also provides guidelines for new constructions. 

\subsection{Connection to Morse functions} Let us start by  making the connection with Morse theory \cite{PMIHES_1988__68__99_0}. 
Suppose $\lambda_1(s)$ and $\lambda_2(s)$ are two parallel planar curves; i.e., their tangent vectors span  a two dimensional vector space  (e.g., two eigenvalues of the time dependent Hamiltonian $H(s)$). Let us assume that $\lambda_1(s)$ and $\lambda_2(s)$ are twice differentiable on a compact interval $I\subset \mathbb [0,1]$ and satisfying $\lambda_1(s)<\lambda_2(s)$ for all $s$ in $I$. Suppose at some point $s^* \in I$, the two curves  converge to each other (without intersecting) and then diverge again; that is, $\lambda_1(s)$ is convex on $I$ while $\lambda_1(s)$ is concave. Without  loss of generality, we can write 
\begin{equation}
\lambda_1(s) = s^2 +\varepsilon \mbox{ and }\lambda_2(s) = -s^2-\varepsilon,
\end{equation}
 with $\varepsilon>0$.
The spectral gap $\Delta(s)=\lambda_2(s)-\lambda_1(s)=2s^2+2\varepsilon$
is minimized at the anti-crossing $s^*=0$; see Figure 1. Now, we embed the two planar curves into the three dimensional Euclidean space 
$\mathbb E^3$ and define the function 
$$
    f(s, \lambda) = (\lambda - \lambda_1(s))  (\lambda - \lambda_2(s)). 
$$
The graph of the function $f$ is the saddle surface (also depicted in Figure 1). With this embedding, the anti-crossing point $s^*$ is mapped into a non degenerate critical point $(s^*, \lambda^*)$ of $f$ with a negative  Gaussian curvature \footnote{Because $f\in \mathbb Q[s, \lambda]$, a point $p$
in $M$ is a non degenerate critical point of the function $f$ if and only if the remainder  ${\sf NF}_\mathcal B (Hessian(f))$  is not zero at $p$, where $\mathcal B$
is a Groebner basis for the ideal generated by the two polynomials $f_s$ and $f_\lambda$.}. 
As we shall show, this embedding also constrains {\em what} can be done to the picture on the right in order to eliminate the anti-crossings and provides  guidelines to {\em how} it can be done. The key constructions are gradient flows \cite{Hurtubise}, the  Euler characteristic, and the Gauss-Bonnet theorem \cite{doCarmo} which distributes the curvature consistently with  the Euler characteristic. 
\begin{figure}[h]
    \centering
    \subfloat  
    {{\includegraphics[width=3cm]{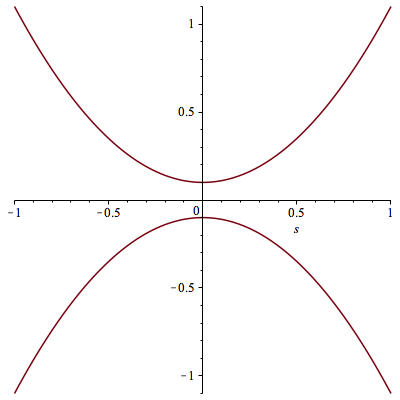} }}%
    \qquad
        \centering
    \subfloat  
    {{\includegraphics[width=3cm]{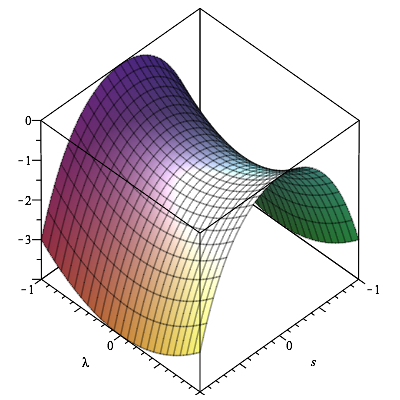} }}%
    \caption{{\small  Anti-crossing happens at saddle points. Note that the picture on the left can be recovered by intersecting the saddle surface (right) with
the horizontal plane $z=0$. The resulting curve is commonly referred to as Dupin indicatrix.}}
    \label{Chimera}
  \end{figure}
  
\subsection{Outline of the paper}
In section  2, we describe the general methodology for changing the dynamics around anti-crossing. We start by reviewing the connection with Morse theory and add more details than the simple connection we made in the Introduction. We then explain that the central object of our investigation
is one (or more) parameter families of Hamiltonians and review the appropriate mathematical machinery, i.e.,  Cerf theory.   The third section discusses
two concrete illustrations of the general theory; these two situations are taken from published literature. The first situation repels the ground state from the rest of the spectrum while the second situation weakens first order quantum phase transitions (QPT) into a higher order. The role of the Gauss-Bonnet theorem is elucidated.   We conclude with some suggestions for future research.

\section{The general theory}
We briefly review the Morse theoretical depiction of the adiabatic evolution of the Hamiltonian (\ref{H}) that we have introduced in \cite{cmu2}. We then extend this depiction to 
parametrized families of Hamiltonians.  For Morse theorists, this extension is referred to as Cerf theory. 
\subsection{Morse theory}
To the Hamiltonian (\ref{H}), we assign the Morse function $f: M\rightarrow [0,1],$ given by the characteristics polynomial (aka secular function)  
\begin{equation}\label{detH}
    f(s, \lambda) = det(H(s)-\lambda I), 
\end{equation}
where $I$ is the identity $n\times n-$matrix. The surface $M$, aka cobordism in the language of Morse theory, is defined by the (non degenerate) critical points
of $f$ using handlebodies decomposition  \cite{cmu2}. This   interplay  between $f$ and the topology (homology) of the cobordism $M$ is the apex of Morse theory, and can be expressed in more precise terms as follows: the trajectories of the gradient flow
 \begin{system}\label{gradientFlowEnergies}
	\frac{d}{d \tau} \lambda(\tau) &=& - \partial_\lambda f (\lambda(\tau) s(\tau)),\\[3mm]
	\frac{d}{d \tau} s(\tau) &=& - \partial_s f (\lambda(\tau), s(\tau)),
\end{system}%
triangulate $M$ and the homology of this complex is isomorphic to the ``usual" singular homology of~$M$.  One can then work on the cobordism $M$ to investigate $f$,
or vice versa. This paper is aligned with the former. 

~~\\
If we Taylor series
$f$ around each of its critical points (and rotate axes to coincide with principal curvature directions), we get a simple local expression of $f$: 
\begin{lemma}[Morse lemma - Revisited]\label{MorseLemma}
Given $p=(s_p, \lambda_p)\in cr(f),$ i.e., a non degenerate critical point of the smooth function $f: M\rightarrow [0,1]$. We have
\begin{equation}\label{MorseEq}
	f(s, \lambda) = f(p) + K_1(p) (s-s_p)^2 + K_2(p) (\lambda-\lambda_p)^2 + higher\, order\, terms,
\end{equation}
with $K_i(p)$ being the principal curvature of $M$ at $p$. 
\end{lemma}
The lemma, which we use repeatedly, gives the two energies around the critical point $p$ as  solutions of the equation (\ref{MorseEq}). In particular, the energy difference between the two
 is given by 
\begin{equation}\label{Delta}
	\Delta(s) = 2\,{\frac {\sqrt {- K_2(p) \left( {\it f(p)}+ K_1(p){s}^{2}
 \right) }}{{K_2(p)}}}
\end{equation}
(after translating $p$ to the origin--it is important to notice that translations, as well as rotations- above,  are isometries, i.e., distance preserving transformations). This gives 
\begin{equation}\label{minDelta}
	\mathrm{min}_s\Delta(s) = 2\,{\frac {\sqrt {- K_2(p) \left( {\it f(p)}
 \right) }}{{K_2(p)}}}.
\end{equation}
The square root is always real-valued: if $p$ is a saddle, i.e., an anti-crossing, then $K_1(p)<0<K_2(p)$ and $f(p)<0$. If
$p$ is a maximum then $K_1(p)<0>K_2(p)$ and $f(p)>0$. If $p$ is a minimum then $K_1(p)>0<K_2(p)$ and $f(p)<0$.

~~\\ Remarkably, the different local descriptions of $f$ can be glued together. The fact that these local depictions can be glued together is the incarnation of the globality of Morse theory. In the language of {\it functors}, this can be presented  as follows: Let $ \mathcal O_M$ denote the category of standard open sets of $M$
sorted by inclusion and $C^\infty(M)$ denote the category of real-valued smooth functions sorted by restrictions. 
\begin{proposition}
The functor $\mathcal F_M:  \mathcal O_M \rightarrow C^\infty(M)$, defined by the lemma for each open set, is a sheaf. 
\end{proposition}
For readers familiar with this language, the proof is almost evident, particularly when  we take into account that critical points are isolated. 

\subsection{Cerf theory}
As mentioned before, the central object in this paper is not the mere Hamiltonian (\ref{H}) but  the one-parameter families of Hamiltonians 
\begin{equation}\label{Hb}
    H^b(s) = H(s) + b H_{enhancement}. 
\end{equation} 
This gives rise to a one-parameter family of Morse functions $f^b: M^b\rightarrow [0,1]$ and to the homotopy:
\begin{equation}\label{h}
h:M^0\times [0, 1] \rightarrow M^1,    
\end{equation}
with $h(s, \lambda, 0) = f^0(s, \lambda)$ 
and $h(s, \lambda, 1) = f^1(s, \lambda)$. 
An important implication of  the homotopy (\ref{h}) is that the  Euler characteristic is given independently of $b$ by the Gauss-Bonnet formula: 
\begin{equation}
    \int_{m \in M^b} K^b d m  = 2\pi \chi. 
\end{equation}
Because the curvature is essentially dumped at the critical points, the above formula constrains  the quantities $\{K^b(p^b), p^b\in cr(f^b)\}$, and,  by (\ref{Delta}), the energy differences, simultaneously.  In fact, for each critical point $p^b\in cr(f^b)$, we have
\begin{equation}
	\mathrm{min}_s\Delta_{p^b}(s) = 2\,{\frac {\sqrt {- K_2^ b(p^b) \left( {\it f^b(p^b)}
 \right) }}{{K_2^ b}}}
\end{equation}
while at the same time we also have
\begin{equation}
	\sum_{p^b\in cr(f^b)}  K_{1}^b(p^b)K_{2}^ b (p^b)  \int _{V^b(p^b)} dm= 2\pi \chi,
\end{equation}
where  the neighbourhoods $V^b(p^b)$ are the graph of $f^b$ (given by the  lemma).  Equivalently, we are constrained (in our quest of enhancement)  by the commutativity of the following diagrams:
 
 \begin{center}
\begin{tikzcd}
f^0 \arrow[to=Z, "{h(- ,b)}"] \arrow[to=2-2]
& {} \\
& \qquad  \qquad    H_*(M^b, \mathbb Z)=H_*(M^0, \mathbb Z)\\
|[alias=Z]|  f^b  \arrow[to=2-2]
\end{tikzcd}
\end{center}
where $f^b\in \Gamma \mathcal{F}_{M_b}$ is the unique global section of the functor $\mathcal{F}_{M_b}$. 
In particular, as $b$ changes, the set of critical points changes but consistently with the invariant homology groups. This means that critical points might move around and  new critical points
might appear (born), or old ones disappear (death), by pairs of (saddle, minimum) or  (saddle, maximum), which amounts to adding, or deleting, a cylinder (surface with $\chi=0$), thus not affecting the topology. 

\section{Two illustrative examples} 
We give two illustrations of the general methodology described above, i.e, homotopically altering the dynamics around the anti-crossing by adding critical points to its vicinity   (through the process of birth/death of critical pairs).  In the first example, we push away the first excited state from the ground state by adding degeneracy, whilst in the second, we change a first order quantum phase transition (QPT) into a second order by adding {\it non-stoquasticity}. The role of the Gauss-Bonnet theorem is also exhibited in both cases. 

\subsection{Enhancement by degeneracy: Repelling the excited state}
This section revisits the work of Dixon \cite{Dickson1}.

\begin{figure}[h]
    \centering
    \subfloat  []
    {{\includegraphics[width=5cm]{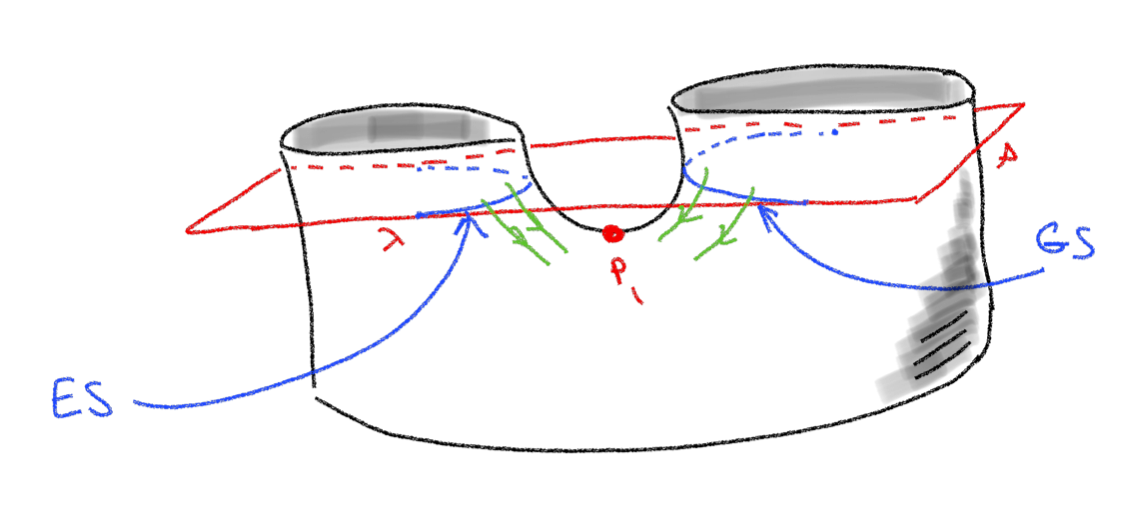} }}%
        \centering
    \subfloat  [ ]
    {{\includegraphics[width=5cm]{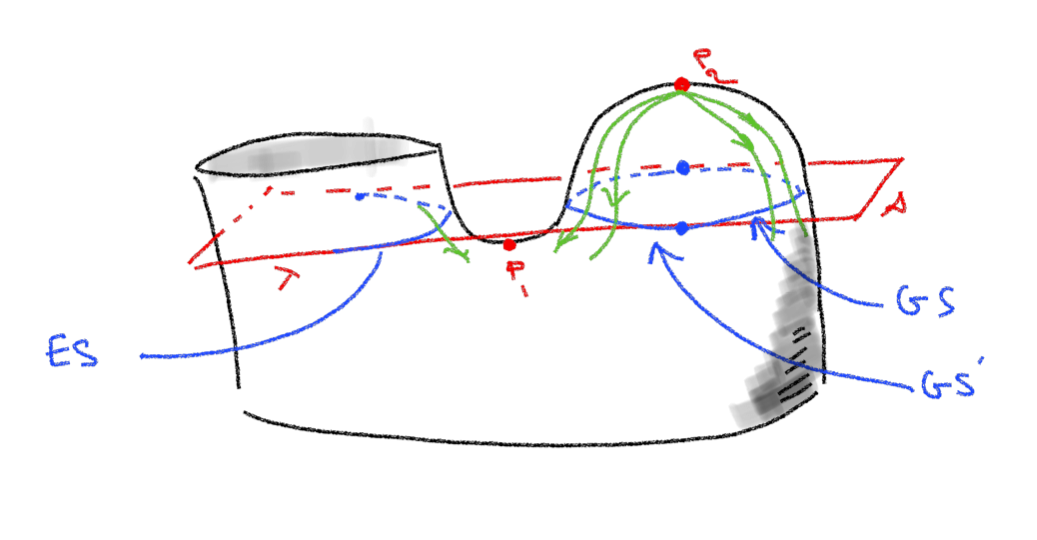} } }%
      \subfloat  [ ]
    {{\includegraphics[width=4cm]{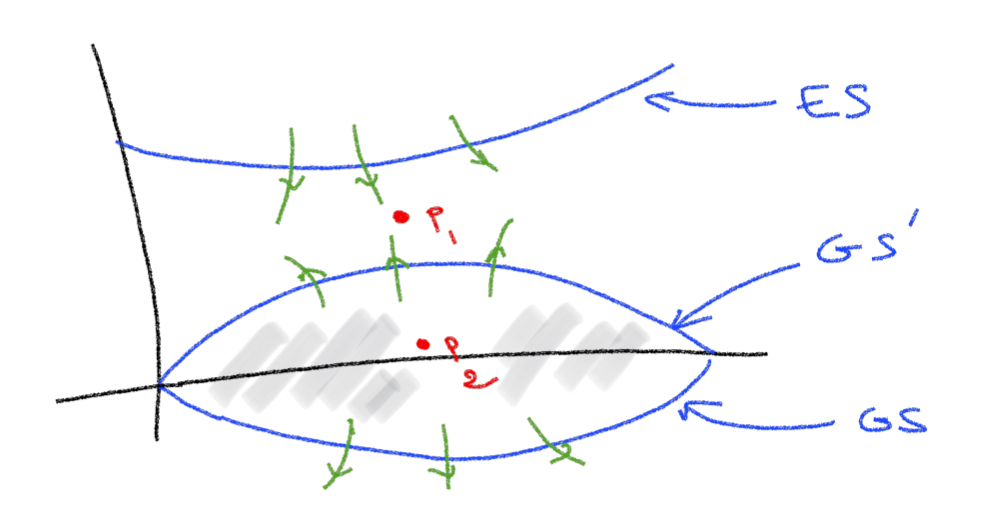} } }%
    \caption{ (a): The initial spectrum lies on  the so-called pair of pants. The two energies are dragged closer by the saddle point i.e., anti-crossing. (b) and (c): 
    The maximum repels the ground state from the rest of the spectrum.  }
    \label{pantsAndCaps}
  \end{figure}

\begin{figure}[h]
    \centering
    \subfloat  []
    {{\includegraphics[width=4cm]{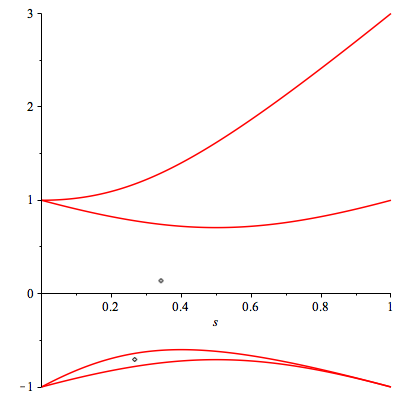} }}%
        \centering
    \subfloat  [ ]
    {{\includegraphics[width=4cm]{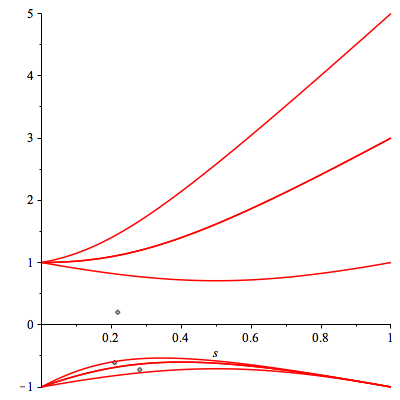} } }%
    \caption{ Two examples that use  the explicit deformation given in  \cite{Dickson1}.  For instance, (b) is obtained by replacing the problem Hamiltonian
    $H_{final} = -\sigma^1$  with $H_{final} + b/2 (\sigma^1+1 ) (1-\sigma^2+1) + b/2 (1-\sigma^1 ) (\sigma^3+1)$. 
      }
    \label{dickson}
  \end{figure}

~\\
The main idea is described in Figure \ref{pantsAndCaps}. Initially, the two lowest energies are attracted to each other due to the presence  of an anti-crossing (as in (a)).
The addition of a minimum  (or maximum as depicted, by making the ground state degenerate at $s=1$, will repel the rest of the spectrum from the ground state, simply because level sets around a minimum  (or maximum) are ellipses ((b) and (c)). See also Figure \ref{dickson}.

~\\   
To make this more concrete,  consider the following situation where we initially have (Morse function of the Grover search adiabatic Hamiltonian \cite{cerf, cmu2})
\begin{equation}
	f(s, \lambda) =\left( {2}^{-N}-1 \right)  \left( {s}^{2} -s\right) +{\lambda}^{2}-
\lambda =
 \left( \lambda- \lambda_2(s) \right) \times \left( \lambda-\lambda_1(s) \right). 
\end{equation}

\begin{figure}[h]
    \centering
    {{\includegraphics[width=2cm]{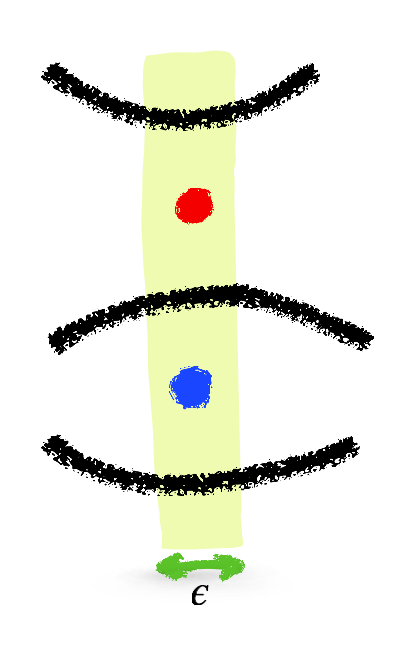} }}%
    \caption{Gauss-Bonnet theorem constrains the energy differences.}
    \label{mexican}
  \end{figure}

~~\\
Similar to \cite{Dickson1}, we make the ground state  degenerate at $s=1$ by defining the one parameter family of functions  ($b>0$): 
\begin{equation}
	f^b(s, \lambda) = \left( \lambda- \lambda_2(s)-b) \right) \times \left( \lambda-\lambda_1(s) \right) \times \left( \lambda-  b) \lambda_1(s) -\lambda_1(s)\right). 
\end{equation}
Here $f^0(s, \lambda) = f(s, \lambda)  \times  \left( \lambda-\lambda_1(s) \right).$ Let us, for instance, fix $b$ to 1 and $N$ to~5.   The Morse function $f^1(s, \lambda)$ now has two critical points in the rectangular region $[1/2-\epsilon/2, 1/2+\epsilon/2] \times [0, R]$ where $R$ is a large number (see Figure \ref{mexican}): 
\begin{equation}
	p^1= (s = 1/2, \lambda = 0.78),\, q^1= (s = 1/2, \lambda = 2.05). 
\end{equation}

\begin{figure}[h]
    \centering
    {{\includegraphics[width=4cm]{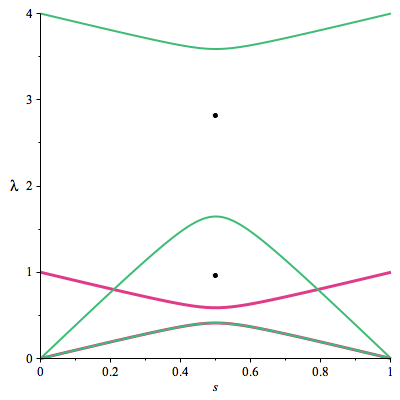} }}%
    \caption{The spectrum of $f^0$ is in purple. The spectrum  of $f^1$ is in green; the ground state is degenerate at $s=1$. Both spectra share the same ground state. The green spectrum  has three energies. The first two energies are separated by a maximum $p_1= (s = 1/2, \lambda = 0.78)$; figuratively, because level sets around $p_1$ are ellipses, the two energies repel each other,  and this keeps the ground state away from the rest of the spectrum. }
    \label{twoSpec}
  \end{figure}

~~\\  
Figures  \ref{mexican} and \ref{twoSpec} also show that $p^1$ is  a maximum while $q^1$ is an anti-crossing point (a saddle). 
When $\epsilon$ is infinitesimally small (it makes sense to take it infinitesimally small, because the addition of $a$ is   affecting ony the $\lambda$ direction; hence, we would like to concentrate on this direction), the two principal curvatures are essentially the same, i.e., $K_2$ (recall that the principal curvatures are by definition the maximum and minimum  of the curvatures of all curves passing through the given point, and because the surface shrinks to a curve, the two are equal).  The application of the Gauss Bonnet formula on  the region $M\cap [1/2-\epsilon/2, 1/2+\epsilon/2] \times [0, R]$,
with $M$ being the surface depicted in (b) above, gives
\begin{equation}
	K_2(p^1)(R-\frac{2}{3}R^3 K_2(p^1) ) + K_2(q^1)(R-\frac{2}{3}c^3K_2(q^1) )  +  O(\epsilon) = 0,
\end{equation}
which gives the functional dependence between $K_2(p^1)$ and $K_2(q^1)$.  
Returning to our instance, a direct calculation of the two principal curvatures confirms this dependence:
$
	 K_2(p_1) = - K_2(q_1) = -3.81. 
$

\subsection{From first order QPT to second order}
When the system endures a first order QPT, the spectral gap between the two lowest energies is exponentially small, yielding exponentially  large evolution time. The reason is quite simple. In  a first order QPT, the two local minima that the ground state occupies before and after the transition are too far apart (in the Hamming distance sense), so the tunnelling is exponentially  slow.   This singularity is expressed mathematically by the discontinuity of the first derivative of  the free energy.  A higher order QPT comes with less severe shrinking of the spectral gap. 

~~\\
The bare bone idea of this section is depicted in the picture below. Initially, we have an anti-crossing in the region $M^0\cap B$. The angle
$\theta $ 
between the two asymptotes is given by 
\begin{equation}
\theta= \pi -\arccos \left( {\frac {K_1+K_2}{K_1-K_2}} \right) = \pi-\arccos \left( {\frac {m}{\sqrt {{m}^{2}-4\,K}}} \right).  
 \end{equation}
Here $m$ is the mean curvature.   For first order QPTs,  this angle is large, i.e., approaches
$\pi$. The figure on the right shows the birth of a pair (min, saddle)  in the vicinity of the anti-crossing.  The effect of this homotopical  change is that the new angle 
$\theta$ is reduced.    As before, the curvatures of the three critical points are related with the relation 
$
	 \int K = -2\pi.
$

 \begin{figure}[h]
    \centering
    \subfloat  []
    {{\includegraphics[width=4cm]{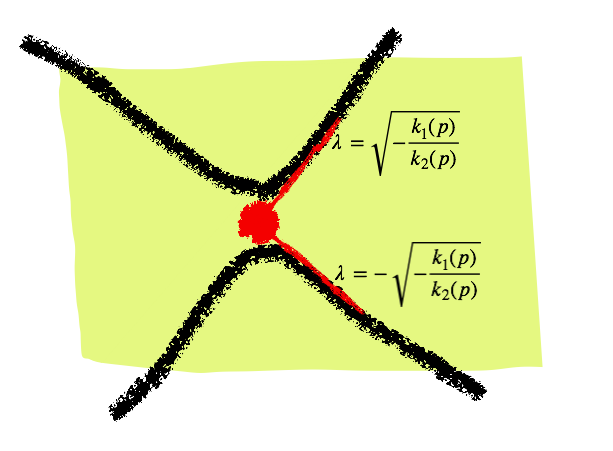} }}%
    \subfloat  []
    {{\includegraphics[width=6cm]{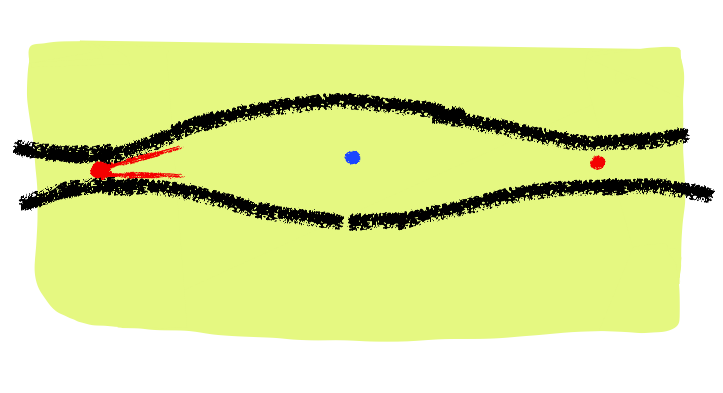} }}%
    \caption{A first order QPT becomes second order by adding the pair (min, saddle). The angle between the two asymptotes, which was initially large (a), is now reduced (b).}
    \label{pattern}
  \end{figure}

~~\\
Let us make this more concrete by considering the following, somewhat simplified, scenario:
 \begin{figure}[h]
    \centering
    {{\includegraphics[width=9cm]{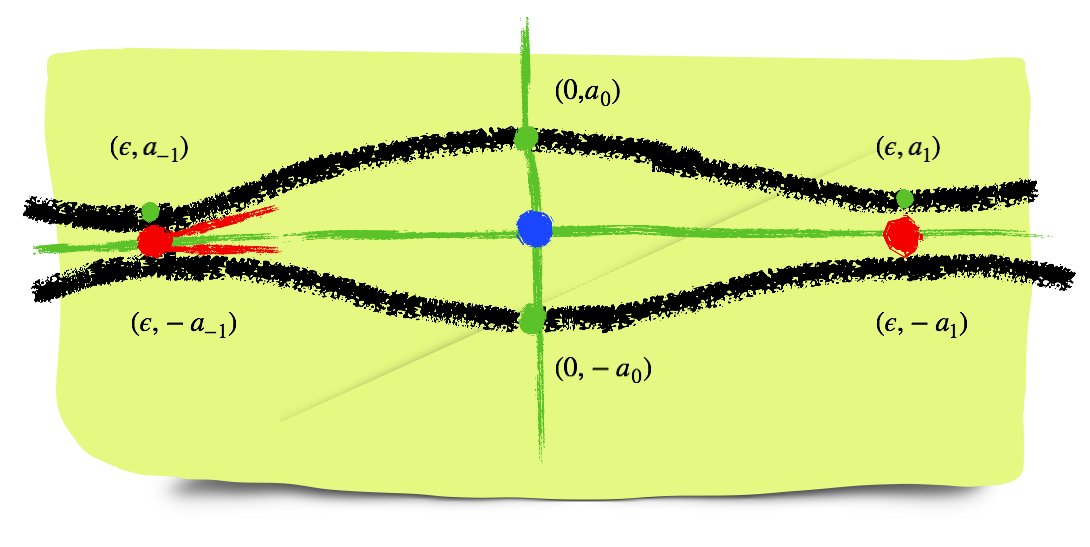} }}%
    \caption{Illustrative scenario.}
    \label{charts}
  \end{figure}  
 ~~\\ 
The Morse function is given by
\begin{equation}
	f^\epsilon(s, \lambda) =  (\lambda-\lambda^*(s)) (\lambda + \lambda^*(s)) 
\end{equation}
with
\begin{equation}
	\lambda^*(s) = \left( \frac{2}{3}\,{\frac {{s}^{2}}{{\epsilon}^{2}}}-\frac{1}{6}\,{\frac {{s}^{4}}{{
\epsilon}^{4}}} \right)  \left( a_{{1}}-2\,a_{{0}}+a_{{-1}} \right) +
 \left( \frac{1}{6}\,{\frac {{s}^{3}}{{\epsilon}^{3}}}-\frac{2}{3}\,{\frac {s}{
\epsilon}} \right)  \left( a_{{-1}}-a_{{1}} \right) +a_{{0}}
\end{equation}
and $a_{i}=1/{K^\epsilon(p_i)}$ for $i=-1,0,1$. Fix $B= [-2\epsilon,2\epsilon] \times [-2a_0, 2a_0]\times [-R, R]$,
with $R$ a large enough positive number-- precisely,  ${R\geq max(\{|f^\epsilon(p_i)|,\, i=-1,0,1  \})}$. Working out
Gauss-Bonnet formula on $M\cap B$ yields the dependancy between the different curvatures:
\begin{eqnarray}\nonumber
	-2\pi &=& {\frac {32}{2835}}\,a_{{0}} \left( -2929\,{a_{{0
}}}^{2}+896\,{a_{{-1}}}^{2}+896\,{a_{{1}}}^{2}-32\,a_{{1}}a_{{0}}-32\,
a_{{0}}a_{{-1}}+256\,a_{{1}}a_{{-1}} \right) \epsilon\\\nonumber
&+& {\frac {16}{2835 
}}\, \left( -10489\,{a_{{0}}}^{2}+896\,{a_{{1}}}^{2}-32\,a_{{1}}a_{{0}
}+256\,a_{{1}}a_{{-1}}-32\,a_{{0}}a_{{-1}}+896\,{a_{{-1}}}^{2}
 \right) \epsilon\,R\\
 & -& \frac{16}{3}\,{a_{{0}}}^{3}R.
\end{eqnarray}
A better example would be  to have two different parameters instead of just $\epsilon$, so the distance between the minimum  and the saddle is independent of their distance from the original saddle. The computation generalizes easily but gives complicated expressions.

\subsubsection{The case of the ferromagnetic $p-$spin model }
\begin{figure}[h]
    \centering
    \subfloat  []
    {{\includegraphics[width=4cm]{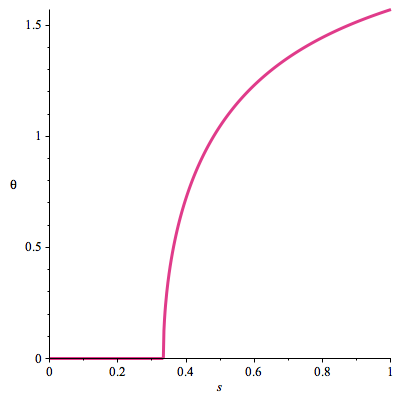} }}%
        \centering
    \subfloat  [ ]
    {{\includegraphics[width=4cm]{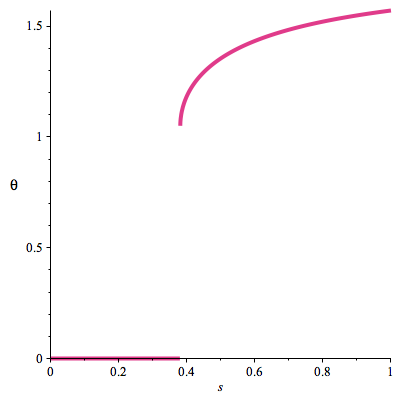} } }%
    \\
      \subfloat  [ ] 
    {{\includegraphics[width=4cm]{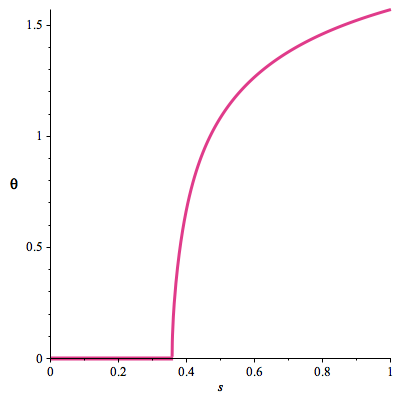} } }%
    \caption{ (a): The minimum of the energy $e(p=2)$ is continuous function of $s$; this minimum is  given by $\theta=\pi -\arccos \left({\frac {-1+s}{2s}} \right)$. The phase transition is second order. (b):
    The minimum of the energy $e(p=5)$ is discontinuous at $s=0.38$. The phase transition is first order. (c) After the addition of the non-stoquastic term with $b =0.1$, the  minimum of the energy is now a continuous function of~$s$. }
    \label{thermo}
  \end{figure}

Here we pay a particular attention to the very instructive model of the  ferromagnetic $p-$spin. See the work by Nishimori and Takada \cite{10.3389/fict.2017.00002}. The time dependent Hamiltonian
is given simply by 
\begin{equation}
	H(s) = -s N\left ( \frac{1}{N} \sum_{i=1}^N \sigma_i^z\right)^p - (1-s) \sum_{i=1}^N \sigma_i^x, 
\end{equation} 
where the $p-$th power refers to the matrix power; i.e., composition of linear operators on the Hilbert space ${\mathbb C^2}^{\otimes N}$. A comprehensive
investigation into the Hamiltonian $H(s)$ can be found in \cite{Bapst_2012}. This Hamiltonian exhibits a first order QPT for $p\geq 3$;  this can be shown in various ways--for instance,  by taking the thermodynamical limit $N=\infty$, in which case one gets  a classical (continuous) spin of a single particle. Figure \ref{thermo} shows
the continuity and discontinuity of the first derivative of the classical energy $e(p,s)$ when $p=2$ and $p=5$ respectively. 

\begin{figure}[h]
    \centering
    {{\includegraphics[width=3cm]{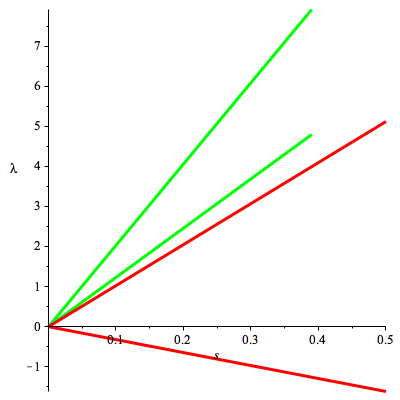} }}%
    \caption{ Decrease of the angle between the two asymptotes around the anti-crossing for $N=7$ and $p=5$. In red, the asymptotes at $b=1$. In green, asymptotes at $b=0.1$.  }
    \label{angles}
  \end{figure}

\begin{figure}[H]
    \centering
    \subfloat  []
    {{\includegraphics[width=11cm]{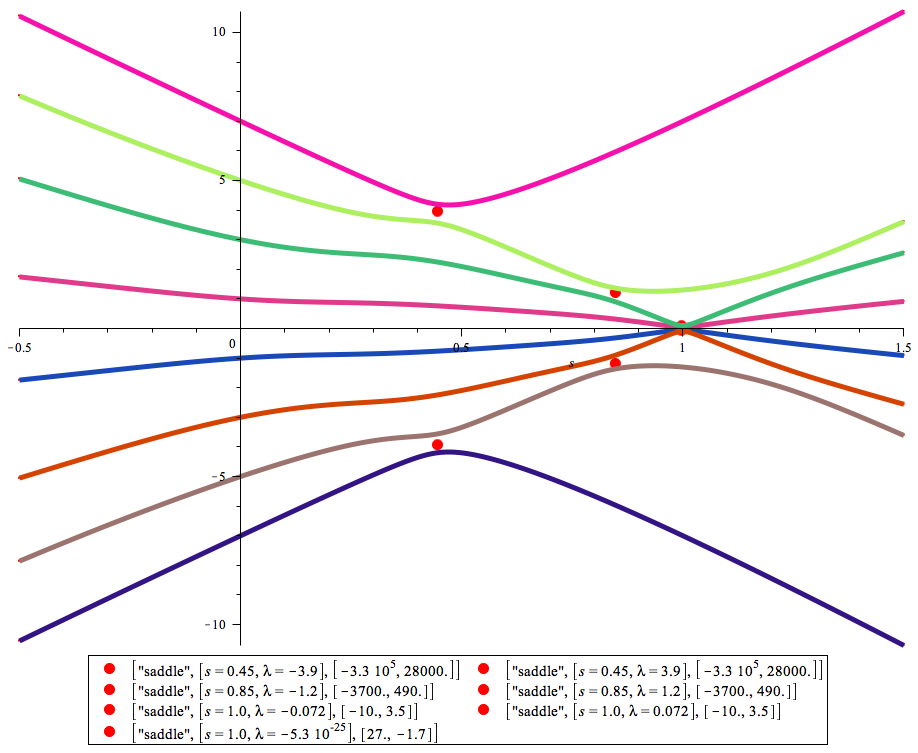} }}%
    \quad 
    \subfloat  []
    {{\includegraphics[width=14cm, height=7cm]{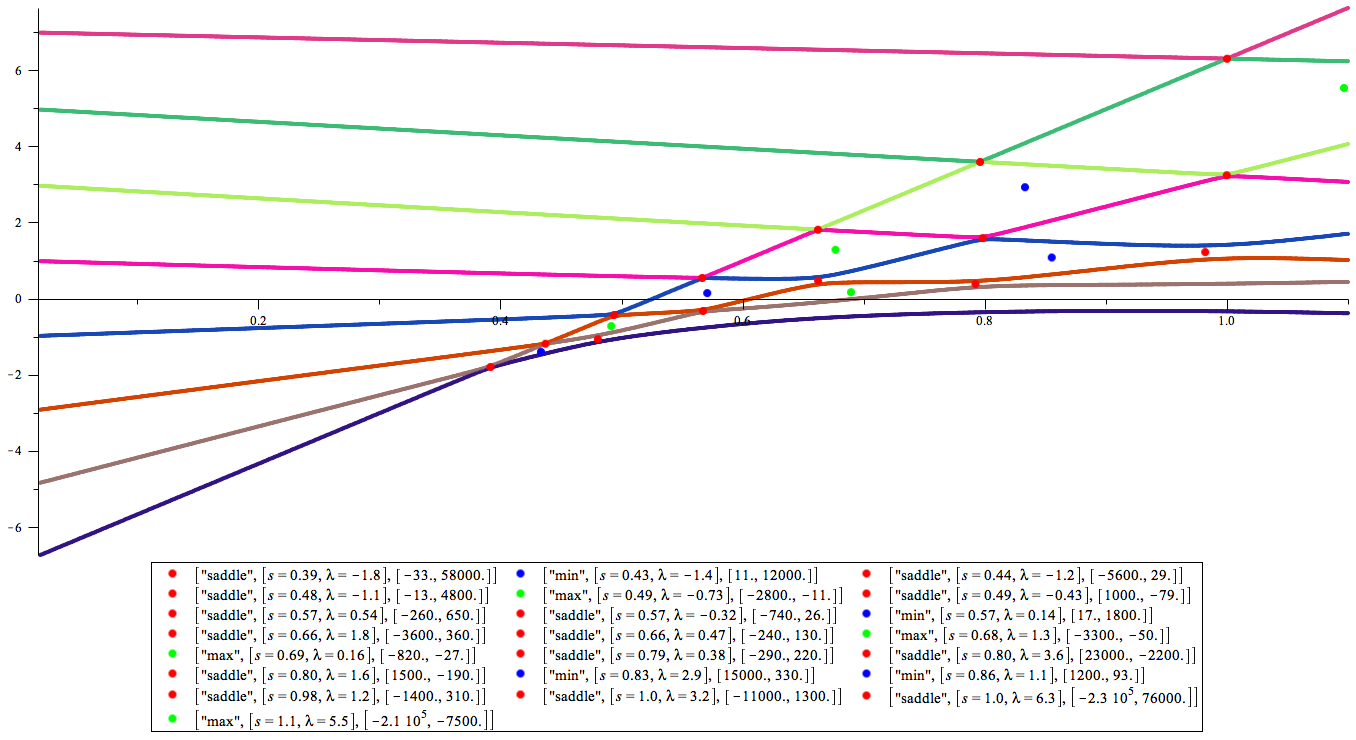} }}%
    \caption{Spectrum of $H(s, b)$ for $N=7$ and $p=5$. (a) $b =1$ (b)  $b =0.1$. The pattern of Figure \ref{pattern} is reproduced with the two lowest energies (at the left most saddle).}
    \label{Nicespec}
  \end{figure}

\begin{figure}[H]
    \centering
    \subfloat  []
    {{\includegraphics[width=8cm, height=6cm]{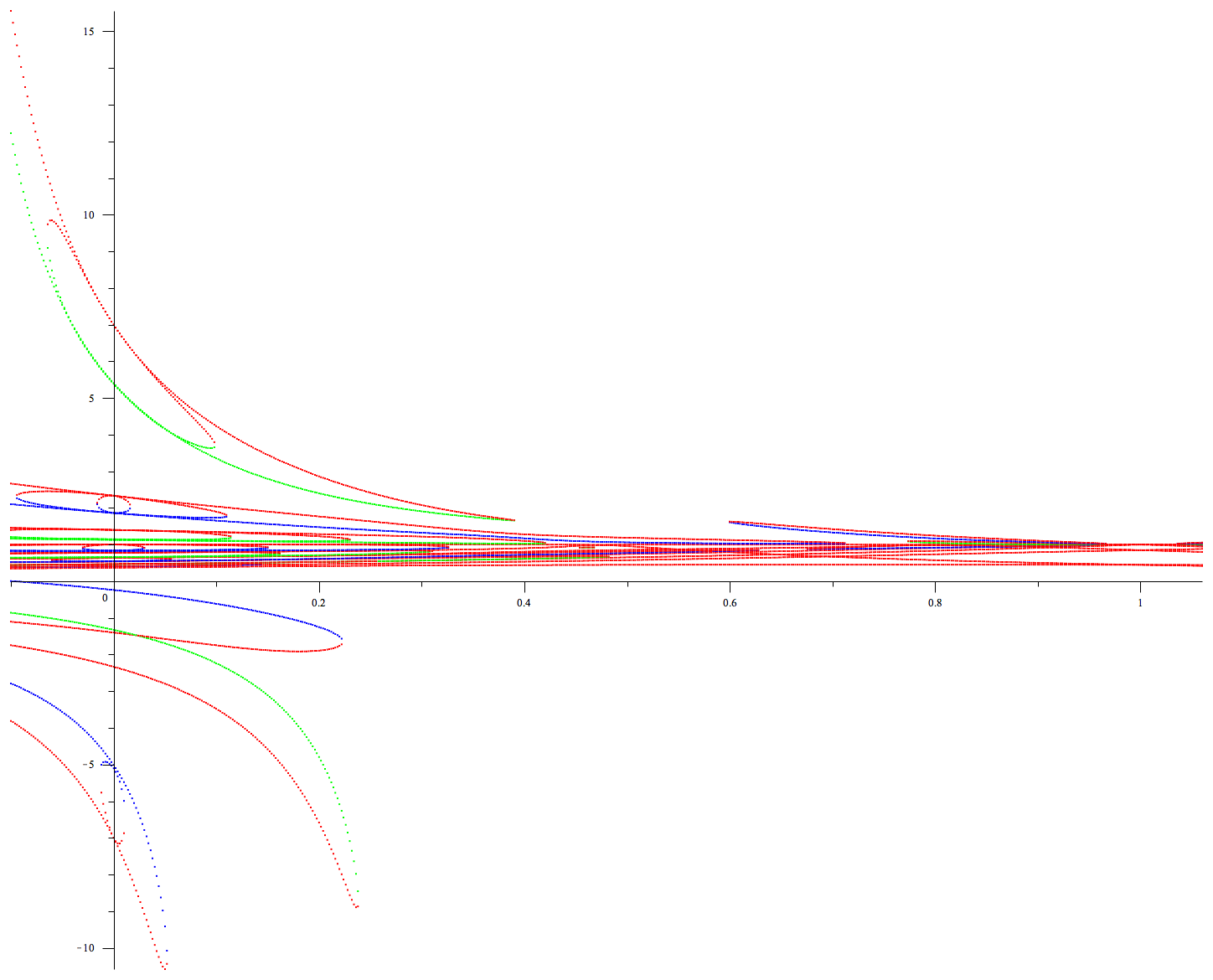} }}%
        \centering
    \subfloat  [ ]
    {{\includegraphics[width=7cm,  height=6cm]{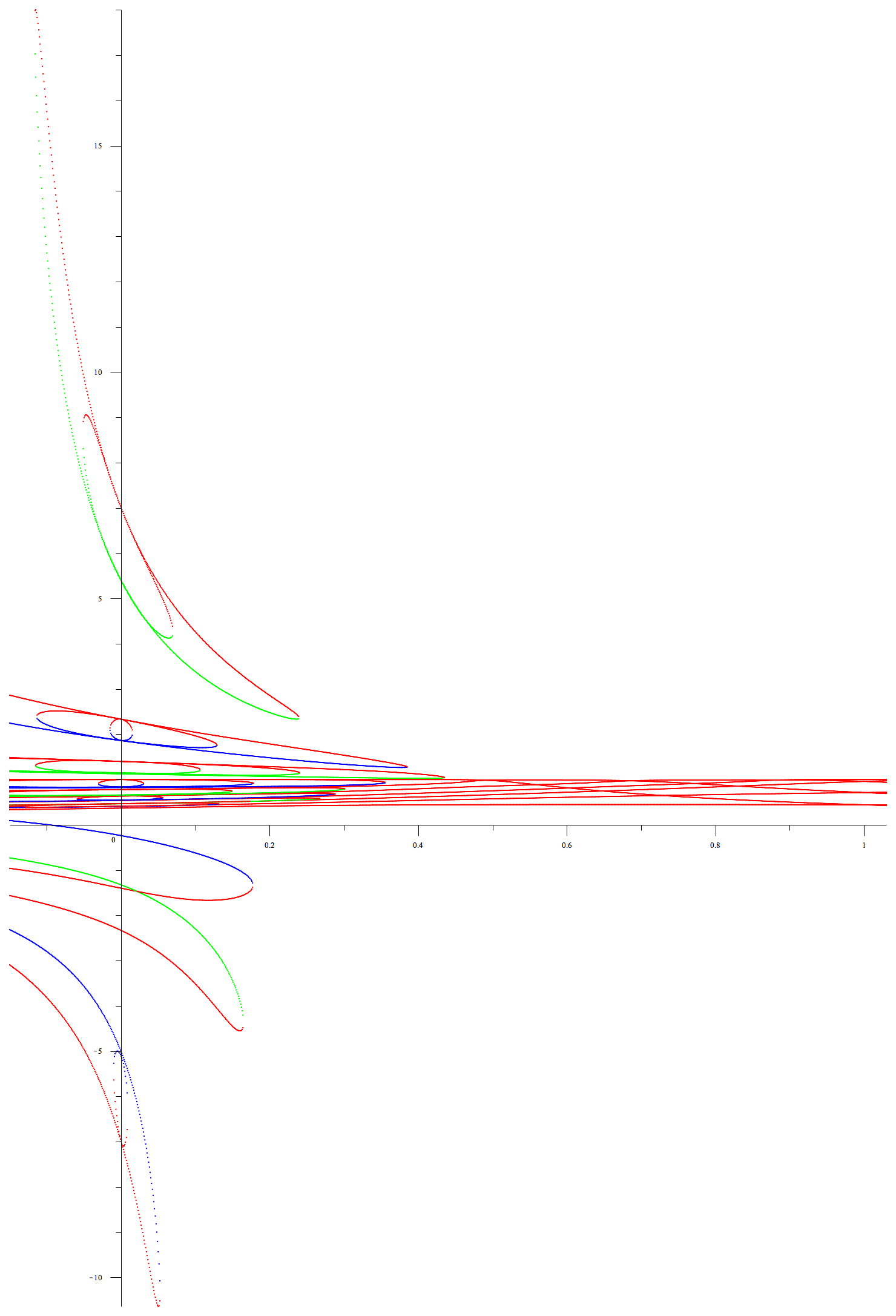} } }%
    ~~\\
      \subfloat  [ ] 
    {{\includegraphics[width=5cm]{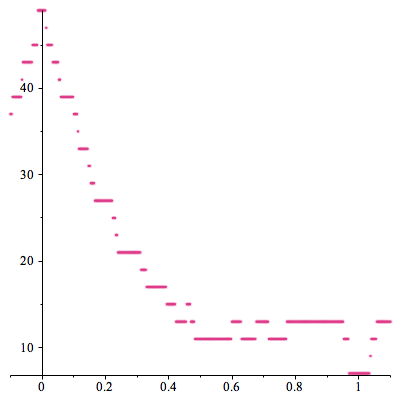} } }%
    \qquad    \quad \quad \, 
      \subfloat  [ ] 
    {{\includegraphics[width=5cm]{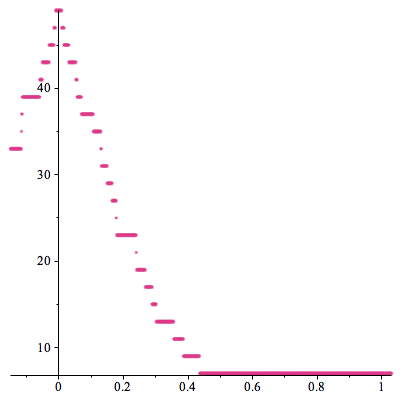} } }%
    \caption{ (a): The evolution of the critical points for $p=5$ and $N=7$. In red are the saddle points (anti-crossings). In green, are the maxima, and in blue, the minima.  Euler characteristic  is constant  $\chi=-7$ for all $b$ in $\mathbb R$. (b): Same for $p=3.$  (c) and(d): 
    The number of critical points (with $N=7$) for $p=5$ and $p=3$ respectively. This number is minimized at $b =1$ and maximized at $b =0$.}
    \label{dance}
  \end{figure}

 ~\\
For finite $N$, the system is also reduced into a much smaller space. This is because of the commutativity of $H(s)$ with the total spin 
$S$. Interestingly, this reduction can be generalized using the theory of irreducible representations \cite{cmurep}, but here we recall the usual reduction.  Conveniently, the  
 lowest part of the spectrum of the total operator is also the lowest part of the spectrum of $H(s=0,b)$. And because of the anti-crossing theorem, this is also true for $H(s, b)$. This lowest part is given by the states with maximum spin $N/2$ i.e.,  states  $|N/2, m\rangle$, for 
 $0\leq m\leq N$,  and defined with 
\begin{eqnarray} 
	S_z |N/2, m\rangle  &=& m |N/2, m\rangle\\ 
	S^2 |N/2, m\rangle  &=& N/2 (N/2+1) |N/2, m\rangle. 
\end{eqnarray}
In addition we have
\begin{equation}
	S_{\pm} |N/2, m\rangle =  \sqrt{N/2(N/2+1) - m(m\pm1)  } \,  |N/2, m \pm 1\rangle
\end{equation}
where $S_{\pm} =S_x \pm S_y$ are the spin raising the lowering operators. All of this  continues to hold with 
the new non-stoquastic Hamiltonian  (\cite{10.3389/fict.2017.00002}): 
\begin{equation}
	H(s, b) =  -s b N\left ( \frac{1}{N} \sum_{i=1}^N \sigma_i^z\right)^p
	+ s(1-b) N  \left( \frac{1}{N} \sum_{i=1}^N \sigma_i^x\right)^k
	 - (1-s) \sum_{i=1}^N \sigma_i^x.  
\end{equation}
A direct calculation gives the matrix elements: 
\begin{eqnarray}\nonumber
	H(s,b)_{m, m} &=&  s \left( -bN \left( 1-2\,{\frac {m-1}{N}} \right) ^{p}+ \left( 1-b
 \right)  \left( 2\,m-1-2\,{\frac { \left( m-1 \right) ^{2}}{N}}
 \right)  \right)
\\[3mm] \nonumber
H(s,b)_{m, m+1} &=& H(s,b)_{m+1, m}  = - \left( 1-s \right) \sqrt { \left( N-m+1 \right) m}\\[3mm] \nonumber
H \left( s,b \right) _{{m,m+2}}&=& H(s,b)_{m+2, m} ={\frac {s \left( 1-b \right) \sqrt {m
 \left( m+1 \right)  \left( N-m+1 \right)  \left( N-m \right) }}{N}}.\nonumber 
\end{eqnarray}
Other matrix elements are zero. 
Sending the non-stoquasticity parameter $b$ to zero amounts to the homotopical reduction of the angle $\theta$ described in the beginning.  This 
pattern   is reproduced in Figure \ref{Nicespec} with the two lowest energies; see also Figure \ref{angles}. Figure \ref{dance} shows how critical points are born and dead 
as $b$ changes without affecting the Euler characteristic. As the parameter  $b$ decreases towards zero, the number of critical points increases, forcing the total curvature to redistribute itself on a larger set of critical 
points.~

%

%
  \newpage
  
  \newpage

%

 \newpage
 \newpage

\section{Conclusion}
The paper presents a general methodology for enhancing the computational power of adiabatic quantum computations. The central idea is to homotopically deform the dynamics around the anti-crossing to weaken its effect.  We have also illustrated the key role the Gauss-Bonnet theorem plays in this process. As illustrated by the two examples, the modifications provide speedup either by repelling the excited state or by transitioning from a first order QPT to the second, and so provide guidance on how to design enhancements. One interesting future direction is to perform such a deformation using degenerate critical points (hence, Conley theory \cite{cmu2}),  as such points are no longer constrained by  the Gauss-Bonnet theorem. This gives more freedom on how to control the evolution.

\bibliographystyle{plain}
\bibliography{c}

\begin{thebibliography}{10}

\bibitem{Hurtubise}
Augustin Banyaga and David Hurtubise.
\newblock {\em Lectures on {M}orse homology}, volume~29 of {\em Kluwer Texts in
  the Mathematical Sciences}.
\newblock Kluwer Academic Publishers Group, Dordrecht, 2004.

\bibitem{Bapst_2012}
Victor Bapst and Guilhem Semerjian.
\newblock On quantum mean-field models and their quantum annealing.
\newblock {\em Journal of Statistical Mechanics: Theory and Experiment},
  2012(06):P06007, jun 2012.

\bibitem{PMIHES_1988__68__99_0}
Raoul Bott.
\newblock Morse theory indomitable.
\newblock {\em Publications Math\'ematiques de l'IH\'ES}, 68:99--114, 1988.

\bibitem{Dickson1}
Neil~G Dickson.
\newblock Elimination of perturbative crossings in adiabatic quantum
  optimization.
\newblock {\em New Journal of Physics}, 13(7):073011, 2011.

\bibitem{doCarmo}
Manfredo~P. do~Carmo.
\newblock {\em Differential geometry of curves and surfaces}.
\newblock Prentice-Hall, Inc., Englewood Cliffs, N.J., 1976.
\newblock Translated from the Portuguese.

\bibitem{cmu2}
Raouf Dridi, Hedayat Alghassi, and Sridhar Tayur.
\newblock Homological description of the quantum adiabatic evolution with a
  view toward quantum computations, {A}rXiv:1811.00675. 2018.

\bibitem{cmurep}
Raouf Dridi, Hedayat Alghassi, and Sridhar Tayur.
\newblock Irreducible representations and {Y}oung diagrams for adiabatic
  quantum computations.
\newblock In preparation. 2019.

\bibitem{Farhi472}
Edward Farhi, Jeffrey Goldstone, Sam Gutmann, Joshua Lapan, Andrew Lundgren,
  and Daniel Preda.
\newblock A quantum adiabatic evolution algorithm applied to random instances
  of an np-complete problem.
\newblock {\em Science}, 292(5516):472--475, 2001.

\bibitem{Hund1927}
F.~Hund.
\newblock Zur deutung der molekelspektren. i.
\newblock {\em Zeitschrift f{\"u}r Physik}, 40(10):742--764, Oct 1927.

\bibitem{10.3389/fict.2017.00002}
Hidetoshi Nishimori and Kabuki Takada.
\newblock Exponential enhancement of the efficiency of quantum annealing by
  non-stoquastic hamiltonians.
\newblock {\em Frontiers in ICT}, 4:2, 2017.

\bibitem{cerf}
J\'er\'emie Roland and Nicolas~J. Cerf.
\newblock Quantum search by local adiabatic evolution.
\newblock {\em Phys. Rev. A}, 65:042308, Mar 2002.

\bibitem{Suzuki2013}
Sei Suzuki, Jun ichi Inoue, and Bikas~K. Chakrabarti.
\newblock {\em Quantum Ising Phases and Transitions in Transverse Ising
  Models}.
\newblock Springer Berlin Heidelberg, 2013.

\bibitem{Vazirani}
Wim van Dam, Michele Mosca, and Umesh Vazirani.
\newblock How powerful is adiabatic quantum computation?
\newblock 2002.

\bibitem{vonNeumann1993}
J.~von Neumann and E.~P. Wigner.
\newblock {\em {\"U}ber das Verhalten von Eigenwerten bei adiabatischen
  Prozessen}, pages 294--297.
\newblock Springer Berlin Heidelberg, Berlin, Heidelberg, 1993.

\end{thebibliography}

\end{document}